\documentclass[aps,prb,twocolumn,10pt,amsmath,amssymb]{revtex4-1}

\usepackage{times}
\usepackage{amsmath}
\usepackage{graphicx}
\usepackage{amsfonts}
\usepackage{amssymb}
\usepackage[utf8]{inputenc}
\usepackage{bm}% bold math
\usepackage{hyperref}% add hypertext capabilities
\usepackage{bbold}
\usepackage{empheq}
\usepackage{mathtools}
\graphicspath{{Figures/}}

\newcommand{\eq}{{\rm eq}}

\newcommand{\vt}{\tilde{v}}
\newcommand{\wt}{\tilde{w}}
\newcommand{\nt}{\tilde{n}}
\newcommand{\pt}{\tilde{p}}
\newcommand{\rt}{\tilde{r}}

\newcommand{\Icalht}{\tilde{\Icalh}}

\newcommand{\Gcalh}{\mathcal{G}}
\newcommand{\G}{\mathcal{G}}

\newcommand{\Sigmacal}{\mathit{\Sigma}}
\newcommand{\Sigmacalh}{\Sigmacal}

\renewcommand{\S}{\Sigmacal}

\newcommand{\HF}{{\rm{HF}}}
\newcommand{\ic}{{\rm ic}}

\newcommand{\Icalh}{{\mathcal{I}}}

\newcommand{\Jcalh}{{\mathcal{J}}}

\newcommand{\Jcalht}{\tilde{\Jcalh}}

\newcommand{\rhob}{\bar{\rho}}
\newcommand{\oneh}{\hat{\mathbb{1}}} %Nice looking 1
\newcommand{\taub}{\bar{\tau}}

\newcommand{\ch}{\hat{c}}
\renewcommand{\dh}{\hat{d}}
\newcommand{\dhd}{\hat{d}^\dagger}

\newcommand{\chd}{\hat{c}^\dagger}

\newcommand{\Hh}{\hat{H}}

\newcommand{\nh}{\hat{n}}
\newcommand{\Nh}{\hat{N}}

\newcommand{\gammah}{\hat{\gamma}}
\newcommand{\dif}{\text{d}}

\newcommand{\Tr}{\text{Tr}}
\newcommand{\trace}[1]{\Tr\left [ #1 \right ]}

\newcommand{\TimeOrdering}[1]{\mathcal{T}\left \{ #1 \right \}}

\newcommand{\Eq}[1]{Eq.~\eqref{#1}}
\newcommand{\Fig}[1]{Fig.~\ref{#1}}

\newcommand{\tb}{\bar{t}}

\renewcommand\Im{\operatorname{Im}}

\begin{document}
\title{The Generalized Kadanoff-Baym Ansatz with Initial Correlations}

\author{Daniel Karlsson}
\affiliation{Department of Physics, Nanoscience Center P.O.Box 35
FI-40014 University of Jyv\"{a}skyl\"{a}, Finland}
\author{Robert van Leeuwen}
\affiliation{Department of Physics, Nanoscience Center P.O.Box 35
FI-40014 University of Jyv\"{a}skyl\"{a}, Finland}

\author{Enrico Perfetto}
\affiliation{CNR-ISM, Division of Ultrafast Processes in Materials
(FLASHit), Area della ricerca di Roma 1, Monterotondo Scalo, Italy}
\affiliation{Dipartimento di Fisica,
Universit\`a di Roma Tor Vergata, Via della Ricerca Scientifica,
00133 Rome, Italy}

\author{Gianluca Stefanucci}
\affiliation{Dipartimento di Fisica, Universit\`a di Roma Tor
Vergata, Via della Ricerca Scientifica, 00133 Rome, Italy}
\affiliation{INFN, Sezione di Roma Tor Vergata, Via della Ricerca
Scientifica 1, 00133 Roma, Italy}

\begin{abstract}
Within the non-equilibrium Green's function (NEGF) formalism, the Generalized Kadanoff-Baym Ansatz (GKBA) has stood out as a computationally cheap method to investigate the dynamics of interacting quantum systems driven out of equilibrium. Current implementations of the NEGF--GKBA, however, suffer from a drawback: real-time simulations require {\em noncorrelated} states as initial states. Consequently, initial correlations must  be built up through an adiabatic switching of the interaction before turning on any external field, a procedure that can be numerically highly expensive. In this work, we extend the NEGF--GKBA to allow for {\em correlated} states as initial states. Our scheme makes it possible to efficiently separate the calculation of the initial state from the real-time simulation, thus paving the way for enlarging the class of systems and external drivings accessible by the already successful NEGF--GKBA. We demonstrate the accuracy of the method and its improved performance  in a model donor-acceptor dyad driven out of equilibrium by an external laser pulse.
\end{abstract}

\maketitle

\section{Introduction}
The real-time Nonequilibrium Green's Function (NEGF) technique~\cite{Danielewicz1984,Stefanucci2013,Balzer2013} for inhomogeneous systems has received a boost in recent years. One of the reasons is the reinvention of the  Generalized Kadanoff-Baym Ansatz (GKBA)~\cite{Lipavsky1986} for the solution of the NEGF equations, which has made it possible to perform \emph{ab-initio} simulations of atoms, molecules, and bulk systems thanks to a drastic reduction of the computational effort. The NEGF--GKBA has been used to study, e.g., atoms~\cite{Perfetto2015b}, biologically relevant molecules~\cite{Perfetto2018}, organic compounds~\cite{Pal2011,Bostrom2018} as well as a large class of extended systems~\cite{Sangalli2015,Sangalli2016} including several two-dimensional layered materials~\cite{Pogna2016,Molina-Sanchez2017}. Recently, the scheme has also been used to study model Hamiltonians with Hubbard or extended Hubbard interactions~\cite{Hermanns2012,Hermanns2013,Hermanns2014,Latini2014,BarLev2016}.

The practical application of the NEGF--GKBA, however, suffers from a drawback. At present it is not known how to include \emph{initial correlations} in the equations of motion;  hence correlations have to be built up in real time. This means taking a noncorrelated state as initial state, evolving the system with an adiabatically switched-on interaction and then continuing the evolution in the presence of time-dependent external fields if nonequilibrium properties are of interest. The NEGF--GKBA formalism, in the most common approximations, contains a memory kernel that makes the computational effort scale quadratically with the number of time steps. Thus, if we need $N_\ic$ time steps to build up initial correlations (using the adiabatic switching) and if the nonequilibrium properties of interest require $N_{\rm prop}$ more time steps, the overall simulation scales like $(N_\ic+N_{\rm prop})^{2}$. Depending on the system $N_\ic$ can be very large, up to the point of making the simulation computationally prohibitive in the physically relevant time window (from $N_\ic$ to $N_\ic+N_{\rm prop}$). Overcoming this drawback would therefore be of utmost practical value.

We stress from the outset that the reduced computational complexity of NEGF--GKBA with respect to NEGF is currently possible only for many-body self-energies up to the second Born (2B) level, with first- and second-order exchange diagrams evaluated using either the bare Coulomb interaction $v$ or the statically or partially dynamically screened interaction $W$. Indeed, the implementation of, e.g., a full GW or T-matrix self-energy would give back the original NEGF scaling in the absence of a GKBA-like expression for the fully dynamically screened interaction $W$ or T-matrix $T$. This current limitation prevents the use of NEGF--GKBA for too strongly correlated systems.

In this work, we extend the NEGF--GKBA equation to allow for starting the real-time evolution from an {\em initially correlated} (IC) state. This allows for driving the system out of equilibrium already at the beginning of the simulation, thereby reducing the scaling of a calculation from $(N_\ic+N_{\rm prop})^{2}$ to $N_{\rm prop}^{2}$. The resulting NEGF--GKBA+IC scheme is general and in principle applicable to any system. Existing NEGF--GKBA codes can easily be extended and the additional computational cost is negligible.

The structure of the paper is as follows. We first give a brief introduction to the NEGF formalism and the GKBA. We then discuss the issue of initial correlations and extend the NEGF--GKBA formalism. Two schemes for calculating the initial correlated state are proposed. We present numerical results in a model donor-acceptor complex, show how our method works in practice and demonstrate its accuracy and improved performance with respect to standard NEGF--GKBA simulations. Finally, we conclude and provide an outlook for future directions.

\section{Kadanoff-Baym Equations}
We consider electrons described by the general time-dependent second-quantized Hamiltonian in a finite basis
\begin{equation}
 \Hh(t) = \sum_{ij\sigma} h_{ij}(t) \chd_{i\sigma} \ch_{j\sigma} +
 \frac{1}{2} \! \sum_{\substack{ijmn\\ \sigma \sigma'}}
v_{ijmn}(t) \chd_{i\sigma} \chd_{j\sigma'} \ch_{m\sigma'}
\ch_{n\sigma}. \label{Hamiltonian}
\end{equation}
The creation (annihilation) operator $\chd_{i\sigma} (\ch_{i\sigma})$ creates (destroys) an electron in basis function $i$ with spin $\sigma$. The single-particle Hamiltonian $h(t)$ contains the kinetic energy as well  as a general time-dependent external field. The two-body interaction $v_{ijmn}(t)$ is taken to be time-dependent in order to describe adiabatic switchings or interaction quenches; we do not specify its specific shape further here. Without any loss of generality we assume that the system is in equilibrium for times $t\leq 0$. For simplicity we consider spin-compensated systems, although no complications arise in the more general case.

We describe the nonequilibrium dynamics of the electrons governed by the Hamiltonian in \Eq{Hamiltonian} using NEGF~\cite{Danielewicz1984,Haug2008,Stefanucci2013,Balzer2013}. The equations of motion for the lesser $\Gcalh^<$ and greater $\Gcalh^>$ single-particle Green's function are known as the Kadanoff-Baym Equations (KBE)~\cite{Kadanoff1962} and read (in matrix form):
\begin{align}
\left [ i \overset{\rightarrow}{\partial}_{t} - h_\HF(t) \right ]& \Gcalh ^\lessgtr(t,t')
\nonumber \\
=&\left [\Sigmacalh^\lessgtr \cdot \Gcalh^A + \Sigmacalh^R \cdot
\Gcalh^\lessgtr + \Sigmacalh^\rceil \star \Gcalh^\lceil
\right](t,t'), \label{KBE1}
\end{align}
\begin{align}
\Gcalh ^\lessgtr(t,t') &\left [ -i \overset{\leftarrow}{\partial}_{t'}
-h_\HF(t') \right ]
\nonumber \\
&\quad\quad=
\left [\Gcalh^\lessgtr \cdot \Sigmacalh^A + \Gcalh^R \cdot
\Sigmacalh^\lessgtr  + \Gcalh^\rceil \star \Sigmacalh^\lceil \right
](t,t'), \label{KBE2}
\end{align}
where we have defined the real-time and imaginary-time convolutions
according to
\begin{align}
\left [A \cdot B\right ] (t,t') &\equiv \int_{0}^\infty \dif \tb\,
A(t,\tb) B(\tb,t'),
\\
\left [ A \star B \right ](t,t') &\equiv -i
\int_{0}^{\beta} \dif \taub A(t,\taub) B(\taub,t'),
\end{align}
with $\beta$ the inverse temperature. The imaginary-time convolutions involve the so-called \emph{mixed} functions with one real time and one imaginary time; they contain information about the IC state~\cite{Stefanucci2013}. The retarded and advanced functions are defined as
\begin{equation}
 X^{R/A}(t,t') = \pm \theta(\pm(t-t')) \left [ X^>(t,t') - X^<(t,t')
\right ]\!. \label{RetAdv}
\end{equation}
The quantity $\Sigmacalh$ in the KBE is the correlation part of the
self-energy. The time-local mean-field or Hartree-Fock (HF) part of
the self-energy  is
incorporated in $h_\HF$, defined as
\begin{equation}
 h_{\HF,ij}(t) = h_{ij}(t) + \sum_{mn} w_{imnj}(t) \rho_{nm}(t),
 \label{HFhamiltonian}
\end{equation}
where $\rho(t)= -i \Gcalh^<(t,t)$ is the single-particle density
matrix and we have defined $w_{imnj}(t) \equiv 2v_{imnj}(t) - v_{imjn}(t)$.

In this work we consider the 2B approximation to the correlation self-energy~\cite{Perfetto2015b}
\begin{equation}
 \Sigmacalh_{ij}^\lessgtr (t,\tb) =
\smashoperator{\sum_{mnpqrs}}v_{irpn}(t) w_{mqsj}(\tb)
\Gcalh_{nm}^\lessgtr(t,\tb) \Gcalh_{pq}^\lessgtr (t,\tb)
\Gcalh_{sr}^\gtrless (\tb,t).
 \label{2ndBornSigmaLesserGreater}
\end{equation}
For future reference, we note that the calculation of the 2B self-energy scales like $N_b^5$ with the number of basis functions $N_b$ and that for any fixed $t$ and $\tb$ it does not scale with the number of time steps $N_t$.

Knowledge of the lesser/greater Green's functions give access to many observables, e.g., density, current density, spectral function, total energy, etc. Unfortunately, the computational effort to solve the KBE is relatively high since these are integro-differential equations for {\em two-time} functions. Using a time-stepping technique the propagation up to $N_t$ time steps scales like $N_t^3$, provided that the calculation of the self-energy does not scale higher than that~\cite{Stan2009}. For the most common approximations used in the literature, i.e., the 2B, GW and  T-matrix approximations, the full solution of the KBE does indeed scale {\em cubically} with $N_t$~\cite{Myohanen2008,Myohanen2009,Friesen2009,PuigvonFriesen2010}. This cubic scaling is what prohibits long time evolutions in many systems.

To reduce the computational effort we reduce the information contained in the unknown functions. Instead of solving the KBE for the Green's function we solve the equation of motion for the single-particle density matrix $\rho(t)$ which is a {\em one-time} function. The equation for $\rho$ can be derived from the KBE by subtracting \Eq{KBE2} from \Eq{KBE1}, and then letting $t' \to t$~\cite{Kadanoff1962,Stefanucci2013}
\begin{equation}
  \partial_t \rho(t) + i \left [ h_\HF(t),  \rho(t) \right ] = -
\left(\Icalh(t)+\Icalh^\ic(t) + {\rm H.c.} \right ),
  \label{rhoEquation}
\end{equation}
where we have defined the {\em collision integral}
\begin{equation}
 \Icalh (t) \! = \!
 \int_{0}^t \! \! \dif \tb \left [\Sigmacalh^>(t,\tb)
\Gcalh^<(\tb,t) - \Sigmacalh^<(t,\tb) \Gcalh^>(\tb,t) \right ]
\label{CollisionIntegral}
\end{equation}
and the {\em IC integral}
\begin{equation}\
\Icalh^\ic(t) = -i
\int_{0}^{\beta} \dif \taub
\Sigmacalh^\rceil(t,\taub)
 \Gcalh^\lceil(\taub,t).
\label{ICIntegral}
\end{equation}
The IC integral $\Icalh^\ic(t)$ depends on $t$ only
through the integrand, whereas the collision integral $\Icalh (t)$
depends on $t$ through both the integrand and the upper integration
limit. Thus, the calculation of the right hand side of
\Eq{rhoEquation} scales linearly with the number of time steps $N_t$.
This implies that the full propagation of the density matrix scales
{\em quadratically} with $N_{t}$, provided that the calculation of
the self-energy does not scale higher than that.

Although the time-stepping technique for  $\rho$ is numerically
cheaper than for the Green's function, Eq.~(\ref{rhoEquation}) suffers from a fundamental problem:
it is not a closed equation for $\rho$. The collision integral $\Icalh(t)$
involves the off-diagonal (in time) $\Gcalh^{\lessgtr}(t,t')$ and the IC integral
contains the mixed functions. In the
next Section we discuss the Generalized Kadanoff-Baym Ansatz (GKBA)
to transform $\Icalh$ into a functional of $\rho$  whereas in
Section~\ref{GKBAICsec} we present the main result of this work,
i.e., a suitable functional form of $\Icalh^\ic$ in terms of $\rho$.

\section{Collision Integral with GKBA}
The GKBA~\cite{Lipavsky1986} is the following {\em ansatz} for the lesser
and greater Green's function (in  matrix form)
\begin{align}
\begin{split}
 \Gcalh^<(t,t') = &-\left [ \Gcalh^R(t,t') \rho(t') - \rho(t)
\Gcalh^A(t,t')\right ],  \\
 \Gcalh^>(t,t') = &\left [ \Gcalh^R(t,t') \rhob(t') - \rhob(t)
\Gcalh^A(t,t')\right ],
\end{split}\label{GKBA}
 \end{align}
where $\rhob(t) \equiv \oneh - \rho(t) = i \Gcalh^>(t,t)$. Of
course,  \Eq{GKBA} alone does not transform $\Icalh$ into a
functional of the density matrix. We also need to specify the
retarded/advanced Green's functions $\Gcalh^{R/A}(t,t')$. These
functions satisfy their own KBE and the computational advantage would
be lost if we had to solve them numerically. For systems where the
average collision time is smaller than the quasi-particle's lifetime
the effect of the correlation self-energy on $\Gcalh^{R/A}(t,t')$ can
be discarded, and we can employ the HF approximation to
$\Gcalh^{R/A}(t,t')$, i.e.
\begin{align}
 \Gcalh^R(t,t') = -i \theta(t-t') &\TimeOrdering{e^{-i \int_{t'}^{t}
h_\HF(\tb) d\tb}}. \label{Gret}
\end{align}

The calculation of the HF $\Gcalh^R(t,t')$ for all $t'<t$ scales linearly in $t$.
We mention that there are also other approximations to
$\Gcalh^R(t,t')$ with the same scaling. They are written in terms of
$\rho$ only and contain correlation effects to some
extent~\cite{Haug1992,Bonitz1999,Arnaud2005,Marini2013,Latini2014}.
The following discussion applies to these approximations as well.

The expression for the retarded Green's functions, \Eq{Gret},
together with  \Eq{GKBA}, define the GKBA.
Since the HF hamiltonian depends only on $\rho$, see \Eq{HFhamiltonian},
the right hand side of \Eq{GKBA} and hence the self-energy  of
\Eq{2ndBornSigmaLesserGreater} are
functionals of $\rho$. Consequently,
the collision integral $\Icalh(t)$, see \Eq{CollisionIntegral},
becomes a history-dependent functional of $\rho(\tb)$ with $\tb \leq t$.

\section{Initial Correlation Integral with GKBA}
\label{GKBAICsec}

\subsection{Drawbacks of a vanishing IC integral}
Without an expression of $\Icalh^\ic$ in terms of $\rho$, the
equation of motion for the density matrix, \Eq{rhoEquation}, cannot
be solved. NEGF-GKBA simulations are usually performed with
$\Icalh^\ic=0$. However, this is justified only provided that the initial state
is noncorrelated. In fact, in the absence of external fields
$\rho(t)=\rho^{\rm eq}$ should be stationary and consequently
$h_\HF(t)=h_\HF^\eq$ is stationary too. If $\Icalh^\ic=0$ then
\Eq{rhoEquation} at
time $t=0$ implies $[\rho^{\rm eq},h_\HF^\eq]=0$
since $\Icalh(0)=0$. Therefore $\rho(t)=\rho^{\rm eq}$ is solution of
\Eq{rhoEquation} with $\Icalh^\ic=0$ only if $\Icalh(t)=0$ for all $t$, i.e.,
only in the absence of correlations. Viceversa,
a correlated density matrix $\rho^{\rm eq}$ does
not commute with $h_\HF^\eq$ and for it to be stationary in the
absence of external fields,  $\Icalh^\ic$
cannot vanish. This is easily seen by taking
again into account that $\Icalh(0)=0$ and hence \Eq{rhoEquation}
at time $t=0$ implies
\begin{equation}
\Icalh^\ic(0)+{\rm H.c.}=-i \left [
h_\HF^\eq,  \rho^\eq \right ].
\label{stateq}
\end{equation}
The common way to circumvent the problem of initially noncorrelated
states consists in starting from
a noncorrelated $\rho(0)=\rho^\eq$ and then build up correlations by
a slow switching-on of the interaction. The drawback of this
procedure is that the correlation build-up time
can be rather long, like in systems with a small
gap between the ground
state and the lowest excited states.
Suppose that we are interested in studying the
nonequilibrium dynamics for $N_{\rm prop}$
time steps and that $N_\ic$ time steps are necessary for the IC
build-up. The computational effort to perform
the $i$-th time step in the physically relevant time-window scales
like $N_\ic+i$ (since $\Icalh$ in \Eq{CollisionIntegral} contains an
integral from time step 0 to time step $N_\ic+i$) and therefore the
cost of the entire simulation
scales like $(N_\ic+N_{\rm prop})^{2}$.

\subsection{Equivalent expression of the IC integral}
Let us now discuss the removal of the adiabatic switching from the numerical procedure. For this purpose, we inevitably need to find an expression
of the IC integral in terms of $\rho$ which satisfies the
{\em stationarity property}
\begin{equation}
\Icalh^{\ic}(0)=\Icalh(t)+\Icalh^{\ic}(t)
\label{statprop}
\end{equation}
for  any $\rho(t)=\rho^\eq$ solution of the stationary equation (\ref{stateq}).
The difficulty in deriving such an expression
stems from the fact that there is no GKBA-like form for the mixed
functions appearing in $\Icalh^{\ic}$, see again \Eq{ICIntegral}.

The solution to the problem is found by rewriting the IC integral in
an equivalent manner. In Appendix~\ref{generalizedFD} we prove a generalized version of the fluctuation-dissipation theorem and use this generalization in
Appendix~\ref{iceqformapp} to show that the IC integral  in
\Eq{ICIntegral} can equivalently be
expressed in terms of real-time Green's functions according
to (see \Eq{equivalentForm})
\begin{equation}
 \!\! \Icalh^\ic (t)  =
 \int _{-\infty} ^{0} \!\!\! \dif \tb \left [
\Sigmacalh^>(t,\tb) \Gcalh^<(\tb,t) - \Sigmacalh^< (t,\tb)
\Gcalh^>(\tb,t) \right ] \! .
\label{CollisionIntegralInit}
\end{equation}
For $t<0$, when the system is in equilibrium,
\Eq{CollisionIntegralInit} follows from the standard fluctuation-dissipation
theorems for $\G$ and $\S$~\cite{Stefanucci2013}.
With the generalized fluctuation-dissipation theorem of
Appendix~\ref{generalizedFD} one
can show that \Eq{CollisionIntegralInit} is also valid out of
equilibrium, i.e., for $t>0$.
We emphasize that the equivalence between
Eqs.~(\ref{CollisionIntegralInit}) and (\ref{ICIntegral}) is an {\em
exact} result, at zero or finite temperature. For notational convenience, we suppress a convergence factor in \Eq{CollisionIntegralInit}, see \Eq{equivalentForm}.

Let us now employ the GKBA approximation to \Eq{CollisionIntegralInit}. The main advantage of \Eq{CollisionIntegralInit} over \Eq{ICIntegral} is that it contains only lesser and greater Green's functions for which a GKBA exists, and we avoid the necessity of constructing a GKBA for the mixed functions. Therefore, \Eq{CollisionIntegralInit} allows us to transform $\Icalh^\ic$ into a functional of $\rho$.

While \Eq{CollisionIntegralInit} is an exact relation, it is not obvious that the application of GKBA to \Eq{CollisionIntegralInit} will yield a solution that satisfies the stationarity property. Let us prove that
the functional $\Icalh^\ic$ indeed fulfills \Eq{statprop}.
For any stationary $\rho$ and in
the absence of external fields $\Gcalh^{R/A}$ is a function of
the time difference only, see \Eq{Gret}. Via the GKBA, \Eq{GKBA}, the same is
true for the lesser and greater Green's functions and hence for the 2B
self-energy of \Eq{2ndBornSigmaLesserGreater}. Renaming the
integration variable in \Eq{CollisionIntegral} and
\Eq{CollisionIntegralInit} according to $\tb' = \tb - t$ we have that
$\Gcalh^{\lessgtr}(t,\tb) = \Gcalh^{\lessgtr}(0,\tb')$ and hence
$\Sigmacalh^{\lessgtr}(t,\tb) = \Sigmacalh^{\lessgtr}(0,\tb')$.
Using \Eq{CollisionIntegral} and
\Eq{CollisionIntegralInit} this
in turn implies that
\begin{align*}
  \Icalh (t) \! +\!  \Icalh^\ic (t) =
 \int_{-\infty}^t \! \! \! \! \dif \tb \left [\Sigmacalh^>(t,\tb)
\Gcalh^<(\tb,t) - \Sigmacalh^<(t,\tb) \Gcalh^>(\tb,t) \right ]
\\
= \int_{-\infty}^0 \! \! \! \!  \dif \tb \left [\Sigmacalh^>(0,\tb)
\Gcalh^<(\tb,0) - \Sigmacalh^<(0,\tb) \Gcalh^>(\tb,0) \right ]
= \Icalh^\ic (0).
\end{align*}
Therefore, a stationary $\rho^\eq$ satisfying \Eq{stateq} yields a stationary right-hand side in \Eq{rhoEquation} also for positive times, in the absence of external fields. This demonstrates the formal usefulness of \Eq{CollisionIntegralInit} in the GKBA context. In the next section we will discuss the practical implications.

\subsection{Practical implementation of the IC integral with GKBA}
\label{practsec}

To make the NEGF--GKBA+IC scheme practical we have to perform the IC
integral from minus infinity to zero analytically for arbitrary time-dependent drivings switched on at $t>0$.
Let us insert the 2B self-energy of \Eq{2ndBornSigmaLesserGreater}
into the expression for $\Icalh^\ic$:
\begin{equation}
    \Icalh^\ic (t)=\Jcalh^{\ic}(t) - \bar{\Jcalh}^{\ic}(t),
\end{equation}
where
\begin{align}
\Jcalh^{\ic}_{ik}(t)=&\sum_{mn pq rs j }
v_{irpn}(t)\;w_{mqsj} \int _{-\infty} ^{0} \!\!\! \dif \tb
\nonumber \\
&\times \,\Gcalh_{nm}^>(t,\tb) \Gcalh_{pq}^> (t,\tb)
\Gcalh_{sr}^< (\tb,t)\Gcalh_{jk}^< (\tb,t)e^{\eta \tb},
\label{jicdef}
\end{align}
and $\bar{\Jcalh}^{\ic}_{ik}(t)$ is defined as in \Eq{jicdef} with
the replacement $\Gcalh^{\lessgtr}\to \Gcalh^{\gtrless}$. We added the convergence factor $e^{\eta \tb}$ [see \Eq{equivalentForm} for details]. In
\Eq{jicdef} we took into account that the tensor $w$ is independent
of time since we assumed that the Hamiltonian is constant at negative
times (for otherwise the system would not be in equilibrium).
The contributions $\Jcalh^\ic$ and  $\bar{\Jcalh}^{\ic}$ have the
same structure; we then  discuss $\Jcalh^\ic$ only. Since  $\tb < 0 <
t$, the GKBA of \Eq{GKBA} yields
\begin{align}
    \begin{split}
\Gcalh^>(t,\tb) &= \Gcalh^R(t,\tb) \rhob(\tb),
\\
\Gcalh^<(\tb,t) &= \rho(\tb) \Gcalh^A(\tb,t).
\end{split}
 \label{GKBA2}
\end{align}
Furthermore, the retarded/advanced Green's functions in the HF approximation, \Eq{Gret}, satisfies the group property
\begin{align}
    \begin{split}
 &\Gcalh^R(t,\tb) = i \Gcalh^R(t,0) \Gcalh^R(0,\tb), \\
 &\Gcalh^A(t,\tb) = -i \Gcalh^A(\tb,0) \Gcalh^A(0,t).
\end{split}
\end{align}
Therefore, we can rewrite the lesser and greater Green's
functions in \Eq{GKBA2} as
\begin{align}
\begin{split}
 \Gcalh^> (t,\tb) &= i\Gcalh^R(t,0) \Gcalh^>(0,\tb), \\
 \Gcalh^<(\tb,t)  &= -i\Gcalh^<(\tb,0) \Gcalh^A(0,t).
\end{split}
\label{GKBAapprox}
\end{align}
As we shall see below, Eqs.~\eqref{GKBAapprox} allow for isolating the $t$-dependence in
$\Jcalh^\ic(t)$  as well as for performing the integral over $\bar{t}$
analytically.

To ease the notation we define the time-dependent tensor
\begin{equation}
 \vt _{irpn} (t) \equiv \sum _{\nt \pt \rt}\, v_{i \rt \pt \nt}(t)\,
\Gcalh^R_{\nt n} (t,0) \Gcalh^R_{\pt p} (t,0) \Gcalh^A_{r \rt} (0,t).
\label{vtilde}
\end{equation}
We also find it useful to define  $\Jcalht^\ic = \Jcalh^\ic(t)
\Gcalh^R(t,0)$ from which we can get back the original
$\Jcalh^\ic(t)$ through $\Jcalh^\ic(t) = \Jcalht^\ic(t) \Gcalh^A
(0,t)$ [we have used that $ \Gcalh^R (t,0) \Gcalh^A (0,t)= \oneh$].
Inserting \Eq{GKBAapprox} into \Eq{jicdef} and taking into account the
above definitions we have
\begin{align}
 \Jcalht^\ic_{ik} (t)  &=
\sum_{mn pq rs j}  \vt_{irpn}(t) w_{mqsj}
\nonumber \\
  \times &\int _{-\infty}^0 \! \! \! \! \! \dif \tb \
\Gcalh^>_{nm}(0,\tb) \Gcalh^>_{pq}(0,\tb) \Gcalh^<_{sr}(\tb,0)
\Gcalh^<_{j k}(\tb,0)e^{\eta \tb}\!.
\quad\label{TimeIntegrationSeparated}
\end{align}
As anticipated the $t-$dependence
has been isolated since it is now contained only  in
the tensor $\vt$.
To perform the integral over $\bar{t}$ we observe that $h_\HF(\bar{t})= h_\HF^\eq$
for all $\bar{t}<0$ and therefore
\begin{equation}
\Gcalh^R(0,\tb) =[\Gcalh^A(\tb,0)]^{\dag}= -i e^{i h_\HF^\eq \tb} .
\label{graeq}
\end{equation}
Let us work in the eigenbasis of $h_\HF^\eq$. In general, this is {\em not} the basis resulting from a pure HF calculation since $\rho^\eq$ and $h_\HF^\eq$ do not commute in the correlated case, see again \Eq{stateq}. Denoting by $\epsilon_n$ the $n$-th eigenvalue of $h_\HF^\eq$, from \Eq{GKBA2} we have
\begin{align}
\begin{split}
    &\Gcalh^>_{nm} (0,\tb) = -ie^{i \epsilon_n \tb}
 \bar{\rho}^\eq_{nm} ,\\
 &\Gcalh^<_{nm} (\tb,0) = i  \rho^\eq_{nm} e^{-i \epsilon_m \tb}.
\end{split}\label{GlesserGreaterSmallT}
\end{align}
Inserting these expressions into \Eq{TimeIntegrationSeparated} and
manipulating  $\bar{\Jcalh}^{\ic}(t)$ in a similar way
we eventually obtain
\begin{equation}
  \Icalh^\ic(t) = \Icalht^\ic(t) \Gcalh^A (0,t),
  \label{Icalh}
\end{equation}
with
\begin{equation}
  \Icalht^\ic_{ik} (t)  =
 i \sum_{n p r}
 \frac{\vt_{irpn}(t) \wt_{nprk}}{\epsilon_r + \epsilon_k - \epsilon_n
- \epsilon_p + i \eta}, \label{Icalht}
\end{equation}
and the tensor $\wt$ defined according to
\begin{equation}
 \wt_{nprk} \! \equiv \!
 \smashoperator{\sum_{mqsj}} w_{mqsj}
 \left (\bar{\rho}^\eq_{nm} \bar{\rho}^\eq_{pq} \rho^\eq_{sr} \rho^\eq_{jk} -
 \rho^\eq_{nm} \rho^\eq_{pq} \bar{\rho}^\eq_{sr} \bar{\rho}^\eq_{jk} \right). \label{wtilde}
\end{equation}
A few remarks are in order:
\\
$(i)$
Equations~(\ref{Icalh},\ref{Icalht}) together with the definitions in
Eqs.~(\ref{vtilde},\ref{wtilde}) allow for including initial
correlations in the
NEGF--GKBA scheme. The resulting
NEGF--GKBA+IC scheme is the main results of this work and it
consists in solving \Eq{rhoEquation} with nonvanishing
collision integral and IC integral. The latter
is functional of the initial correlated
equilibrium density matrix $\rho^\eq$ and of its time-dependent value
$\rho(t)$ (through the retarded/advanced Green's functions).
\\
$(ii)$
The Coulomb tensor $v$ and hence $w$ are written in the
eigenbasis of $h_\HF^\eq$. Thus, interactions that are sparse in some
basis, such as the Hubbard interaction in the site basis,
do not necessarily yield a sparse tensor $v$ in the eigenbasis of $h_\HF^\eq$.
\\
$(iii)$
In the noncorrelated case $\rho^\eq$ is diagonal and
it is easy to show that the tensor $\wt$ is identically zero for $\epsilon_r + \epsilon_k -
\epsilon_n - \epsilon_p=0$. For a general correlated density matrix $\wt_{nprk}$
vanishes whenever $r=n$ and  $k=p$ or $r=p$ and $k=n$.
We assume the same behavior even for accidental degeneracies
 and restrict the
summation in \Eq{Icalht} to include only those indices for which  the
denominator is non-vanishing. Thus, we can safely set $\eta=0$.
\\
$(iv)$
The extra computational effort for the implementation of the IC
integral is minimal. The calculation of $\wt_{nprk}$ has to be done only once and the summation can be performed efficiently in sequence, scaling at most like $N_b^5$ where $N_b$ is the number of basis functions. The same efficient summation can be used to calculate $\vt_{nprk}(t)$ in \Eq{vtilde}, although in this case the summation has to
be performed for every time step. Having $\wt$ and $\vt(t)$ we calculate  $\Icalht^\ic(t)$ from \Eq{Icalht}, another operation that scales like $N_b^5$.
The scaling with the fifth power of $N_{b}$ is the same as that of the summation involved in the 2B self-energy of \Eq{2ndBornSigmaLesserGreater}. Thus, $\Icalh(t)$
and $\Icalh^{\ic}(t)$ scale in the same way with the number of basis functions. However, the IC integral does not scale with the number of time steps $N_{t}$ (no time integration) whereas the collision integral scales linearly with $N_{t}$ (integration from time step 0 to time step $N_{t}$). Consequently, the inclusion of initial correlations via $\Icalh^\ic(t)$ adds a negligible computational cost to
standard GKBA simulations. Furthermore, the calculation of $\Icalh^\ic(t)$ is completely independent from $\Icalh$ and can be done separately; hence no internal modifications need to be made to an existing GKBA code in order to incorporate initial correlations.
\\
($v$) In Appendix~\ref{appendixA} we show that the above conclusions
remain intact when using a given dynamically screened interaction
$W(t-t')$, as that of Ref.~\cite{Pal2009,Pal2011}, in place of the bare time-local interaction $v$.

\section{The equilibrium correlated density matrix}
\label{eqmethods}

In the NEGF-GKBA+IC scheme the initial and correlated density matrix
$\rho^{\rm eq}$ satisfies \Eq{stateq}, and $\rho(t)=\rho^{\rm eq}$ is a
solution of the equation of motion (\ref{rhoEquation}) in the absence
of external fields. A scheme to obtain $\rho^{\rm eq}$ based on
solving the equilibrium KBE for the lesser Green's function
using the GKBA for the collision integral has
recently been proposed in Ref.~\cite{Hopjan2018}.
In the following we discuss two alternative methods.

The first method consists in solving \Eq{stateq} self-consistently. This equation, however, admits infinitely many solutions since the diagonal of the left and right hand sides vanish in any real basis for Hamiltonians invariant under time-reversal.
In fact, \Eq{stateq} is not a variational equation, rather it is a stationary equation, i.e., it stems from setting $\partial_{t}\rho=0$. The possible solutions do therefore correspond to the infinitely many stationary density matrices of the system. A unique solution can be found by supplementing \Eq{stateq}  with the value of the diagonal occupations $\rho_{nn}=\{f_{n}\}$ in some basis. To illustrate the self-consistent procedure let us first discuss the noncorrelated case, i.e., $\Icalh^{\ic}=0$. Then,  \Eq{stateq} tells us that $\rho^{\rm eq}$ is diagonal in the eigenbasis of $h_{\rm HF}$.
We then diagonalize the noninteracting Hamiltonian $h$,
find the eigenvectors $\varphi^{(0)}_{n}$,
and construct $\rho^{(0)}_{nm}=\delta_{nm}f_{n}$ in the basis of
these eigenvectors. In the $(i+1)$-th iteration step we use
$\rho^{(i)}$ to calculate $h_\HF^{(i)}=h_\HF[\rho^{(i)}]$, find the
eigenvectors  $\varphi^{(i+1)}_{n}$
and construct $\rho^{(i+1)}_{nm}=\delta_{nm}f_{n}$ in the
$(i+1)$-basis. At convergence we have the HF basis with HF occupations
$\{f_{n}\}$. In particular, if $f_{n}=1$ for $n\leq N_{\rm el}$ and
zero otherwise the procedure converges to the HF ground state with
$2N_{\rm el}$ electrons. In the correlated case the procedure is
identical but in the $(i+1)$-th iteration step $\rho^{(i+1)}_{nm}$ is
not diagonal. In the eigenbasis $\varphi^{(i+1)}_{n}$ of
$h_\HF^{(i)}=h_\HF[\rho^{(i)}]$ with
eigenvalues $\epsilon^{(i+1)}_{n}$ we have for $n\neq
m$
\begin{equation}
\rho^{(i+1)}_{nm}=i\;\frac{\Icalh^{\ic}_{nm}(0)+\Icalh^{\ic*}_{mn}(0)}
{\epsilon^{(i+1)}_{n}-\epsilon^{(i+1)}_{m}}.
\label{offdiagrho}
\end{equation}
As already observed this result does not allow to update the diagonal
elements. We could either supplement \Eq{offdiagrho} with
$\rho^{(i+1)}_{nn}=f_{n}$ for some reasonable set of occupations or
take advantage from a self-consistent Matsubara Green's function
calculation providing $\rho_{pq}=\delta_{pq}f_{q}$ in the natural
orbital basis $\psi_{q}$ and supplement \Eq{offdiagrho} with
\begin{equation}
\rho^{(i+1)}_{nn}=\sum_{q}f_{q}|\langle\psi_{q}|\varphi^{(i+1)}_{n}\rangle|^{2}.
\end{equation}
Independently of the prescription to fix the diagonal elements
$\rho^{(i+1)}_{nn}$, at convergence $\rho^{\rm eq}$ satisfies \Eq{stateq}.

The second method is instead borrowed from standard NEGF--GKBA simulations.
We start from an noncorrelated density matrix at time $t=0$ and evolve
the system with no external fields in the presence of a slowly
increasing interaction $v(t)$ having the property that $v(t<0)=0$ and
$v(t>T_\ic)=v$. The time $T_\ic$ is the IC build-up time which should
be chosen large enough for $\rho(t)=\rho(T_\ic)$ to be sufficiently
stationary when $t$ is larger than $T_\ic$. Taking advantage of the
fact that $v(t)=0$ for $t\leq
0$, the IC integral vanishes at all times $t$ since
$\Sigmacalh^\lessgtr (t,\tb)=0$ for $\tb\leq0$, as can be seen from
\Eq{2ndBornSigmaLesserGreater} and \Eq{CollisionIntegralInit}. At the
steady state $\rho(T_\ic)=\rho^\eq$ satisfies \Eq{stateq}. We
emphasize again that the number of time steps for the IC build-up
does not affect the computational cost of the subsequent physically
relevant time propagation with $\rho(0)=\rho^\eq$ as initial
state. We also observe that this second method is limited to systems
at zero temperature. In fact, due to correlation-induced level
crossings and/or splittings of degenerate many-body states, the
finite-temperature noninteracting density matrix does not, in
general, evolve into the finite-temperature interacting
one.

\section{Example of GKBA with initial correlations}
In this section we provide numerical evidence that our procedure works and is efficient. As a non-trivial example, we consider the donor-acceptor dyad used in Ref.~\cite{Latini2014} as a molecular junction to address the ultrafast charge dynamics at the donor-acceptor interface. The system is modelled by a two-levels donor, the levels being the HOMO ($H$) and LUMO ($L$), and a linear chain of $N_{a}$ acceptor sites labelled by the site index $a$. The Hamiltonian reads
\begin{align*}
\begin{split}
\Hh &=  \epsilon_A\sum_{a=1}^{N_{a}}\nh_a
+T_{DA}\sum_\sigma \left ( \chd_{L\sigma} \ch_{1\sigma}
+ {\rm H.c.}\right)
\\
&+\sum_{i=H,L}\epsilon_i \nh_i
+T_A  \sum_{\sigma,a=1}^{N_{a}-1} \left (
\chd_{a\sigma}\ch_{a+1,\sigma} + {\rm H.c.}
 \right)
\\
&+
U_{DA}(t) ( \nh_H \!+\! \nh_L \!-\! 2)
\sum_{a=1}^{N_{a}} \frac{\nh_a- 1}{a},
\end{split}
\end{align*}
where ${\rm H.c.}$ denotes the hermitian conjugates. We defined
$\nh_i = \sum_\sigma \nh_{i\sigma}$ the occupation of level $i=H,L$
with energy $\epsilon_i$
and likewise for the occupation of the acceptor sites. The system is
isolated and the dimension of the single-particle basis is
$N_b=2+N_{a}$.
The LUMO is not coupled to the HOMO but
to the first site of the acceptor chain with tunneling amplitude
$T_{DA}$. The tunneling amplitude between two nearest neighbour
acceptor sites is $T_{A}$.
In accordance with Ref.~\cite{Latini2014} we set the level energies $\epsilon_H =
-2.92$, $\epsilon_L = -0.92$ and $\epsilon_A = -2.08$, and the
tunneling amplitudes  $T_{DA} =
-0.3$ and $T_A = -0.2$ (all quantities are in atomic units). The donor-acceptor dyad is half-filled with equal number of up
and down electrons. The electrons interact  with a
density-density type  of interaction, and we set the interaction
strength $U_{DA}(t) = U_{DA}=0.5$ for positive times.

As time-dependent perturbation we choose
\begin{equation}
 \Hh_{\text{ext}} (t) = f(t)\sum_{\sigma}
 \left ( D_{LH} e^{i \Omega t}
\chd_{H\sigma} \ch_{L\sigma} + {\rm H.c.} \right )
\label{ExtPerturbation}
\end{equation}
describing the coupling between a monochromatic electric field of
amplitude $f$ and frequency $\Omega$, and the HOMO-LUMO dipole moment $D_{LH}$. We consider a
resonant frequency  $\Omega = \epsilon_L - \epsilon_H =
2$ and set the value of $ D_{HL} = 0.3$. The electric field is
very strong, $f=1$, and it is
active from time $t=0$ until time $t= \frac{\pi}{4 D_{LH}}\simeq 2.6$.
As we shall see, the external driving transfer one
unit of electric charge from the initially filled HOMO to the
initially empty LUMO. In all simulations below we have considered the
number of acceptor sites $N_{a}=4$.

\subsection{Simulations without external field}
We first show calculations without external fields to illustrate that
the system is stationary with the inclusion of the IC integral.
We use the adiabatic switching method
to obtain the initially correlated density matrix $\rho^\eq$, see
Section~\ref{eqmethods}.
The switching protocol was chosen to be
\begin{equation}
U_{DA}(t) = U_{DA}
\times\left\{
\begin{array}{ll}
\sin^2 \left ( \frac{\pi}{2} \frac{t}{T_\ic} \right )
& t<T_\ic \\
1 & t\geq T_\ic
\end{array}
\right.
\end{equation}
We have used the CHEERS code~\cite{CHEERS} with time step $\Delta t=0.005$ to
perform three separate calculations: (a) A
calculation with $\Icalh_\ic(t)=0$ that starts from $t=0$ with the noncorrelated
HF density matrix and adiabatically switches on the interaction
 with $T_\ic=100$ (for this calculation we have shifted the time axis
to set the time origin at  $T_\ic$); (b) A NEGF--GKBA+IC calculation  with the IC
integral evaluated as described in Section~\ref{practsec} that starts from
$t=0$ using $\rho(t=0) = \rho^\eq$;  (c) A calculation with $\Icalh_\ic(t)=0$
that starts from $t=0$ using $\rho(t=0) = \rho^\eq$.
We remind that $\rho^\eq = \rho^\eq(T_\ic)$ and hence
calculations (a) and (b) are expected to coincide for
large enough $T_\ic$. We also stress that the computational time for
calculations (b) and (c) is practically equal.

\begin{figure*}[ht]
\begin{center}
 \includegraphics[width=0.78\textwidth]{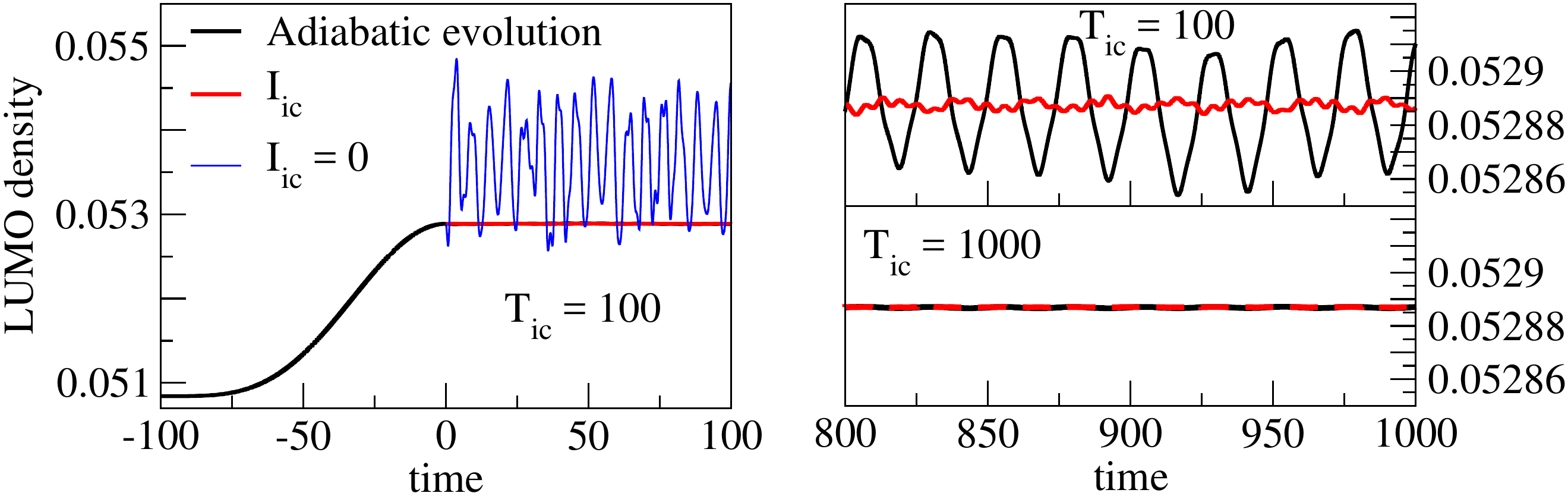}
\end{center}
 \caption{LUMO occupation, without external fields, for the three
type of  calculations described in the main text. Total number of time steps
$N_t=2 \cdot 10^5$ and time step $\Delta t=0.005$.
Left panel: evolution using $T_\ic=100$ up to $t=100$. Top right panel:
long time
behavior for $T_\ic = 100$. Bottom right panel: Long time behavior for $T_\ic =
1000$. In the right panels we do not show the curve corresponding to
calculation (c) [$\Icalh_\ic(t)=0$, see main text]  due to too large oscillations.
 \label{noField1000}}
\end{figure*}
In \Fig{noField1000} we show
the evolution of the LUMO occupation $n_L = \rho_{LL}(t)$ up to
$t=1000$.
From the left panel we conclude that the adiabatic evolution,
calculation (a), yields a LUMO occupation that
remains stationary after $t>0$, except for small oscillations due to
the finiteness of $T_\ic$. The same quantity for
calculation (b), that includes
$\Icalh^\ic$, is indeed stationary, even for very long propagation
times. Calculation (c), where $\Icalh^\ic$ is artificially set to
zero, does instead yield a nonstationary $\rho(t)$, as expected from
the discussion of the previous
Section. For long times, the LUMO occupation for both calculations
(a) and (b)
 (top right panel) shows small
oscillations due to the finite adiabatic switching time. Increasing
the switching time to $T_\ic=1000$ the amplitude of the oscillations
decreases for
both calculations  (bottom right panel in \Fig{noField1000}).
Perhaps remarkably, the correlated density matrix $\rho^\eq$ resulting
from the adiabatic switching with
$T_{\rm ic}=100$ yields a reasonably stationary $\rho(t)$ in
NEGF--GKBA+IC [certainly less oscillatory than that of calculation (a)],
indicating that the NEGF--GKBA+IC equation is numerically stable.

\subsection{Simulations with external field}
\begin{figure*}
\begin{center}
 \includegraphics[width=0.78\textwidth]{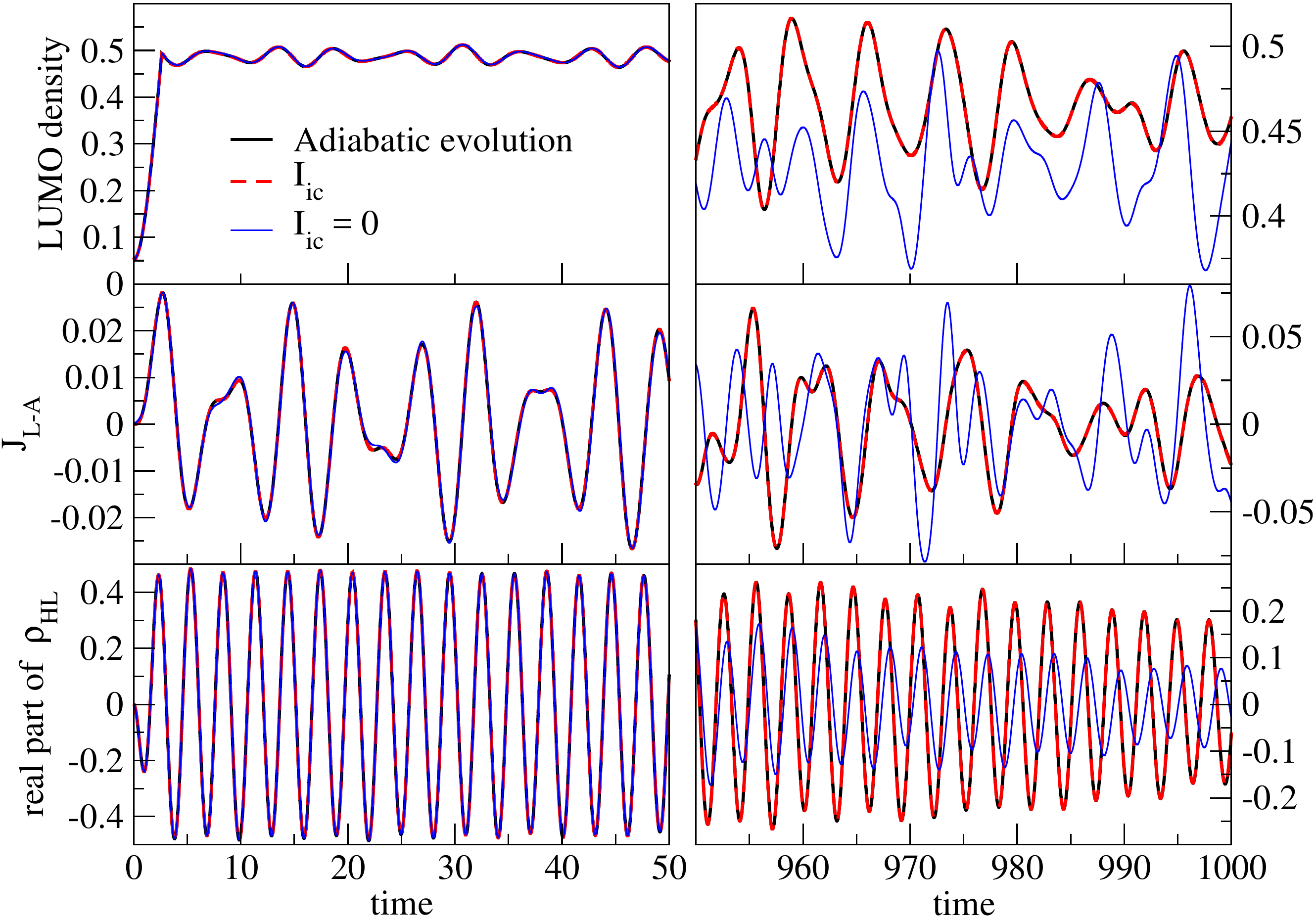}
\end{center}
 \caption{LUMO occupation (top panels), current between LUMO and the first acceptor site  (middle panels) and real part of $\rho_{HL}$ (bottom panels) in the presence of the external driving in \Eq{ExtPerturbation} for the three type of  calculations described in the main text. Total number of time steps $N_t=2 \cdot 10^5$ and time step $\Delta t=0.005$. The quantities are shown in the time range $(0,50)$ (left) and $(950,1000)$ (right).
 \label{all}}
\end{figure*}

We now show that also the off-diagonal elements of the density matrix are well-reproduced in NEGF--GKBA+IC. We perform the three type of calculations of the previous section in the presence of the external driving in \Eq{ExtPerturbation}, and use a very long adiabatic switch-on time $T_\ic=1000$ to converge the calculations. The quantities chosen to illustrate the performance of  the NEGF--GKBA+IC scheme are the LUMO density, the current $J(t) = 2 |T_{DA}| \Im [ \rho_{L1}(t)]$ flowing through the bond between the LUMO and the first acceptor site and the real part of the off-diagonal HOMO-LUMO matrix element of $\rho(t)$. The results are shown in \Fig{all} up to $t = 1000$.

As anticipated the NEGF--GKBA+IC scheme, calculation
(b), correctly reproduces the outcome of standard NEGF--GKBA with
an adiabatically switched-on interaction, calculation (a). The agreement is
excellent all the way to the end of the simulation time.
Neglecting the IC integral and starting from the
correlated density matrix $\rho^{\eq}$, calculation (c),
introduces an error that becomes more severe as the time increases. The general
trend is that all quantities can be well-reproduced for short times even
without properly accounting for initial correlations, but eventually
the agreement  tend to deteriorate.

\section{Conclusions}
Using the NEGF--GKBA+IC scheme we have shown how to separate the calculation of the
correlated density matrix from that of the time-dependent responses.
By generalizing the fluctuation--dissipation theorem for the Green's function and
self-energy we have derived an
equivalent expression of the IC integral suited to be
evaluated using the GKBA.
With the addition of this IC integral
it is possible to use
correlated states as initial states, thus
removing the bottleneck of a preliminary adiabatic switching.
For the most common approximations the computational
effort of our method scales favorably and, most importantly, does not slow down an
ordinary NEGF--GKBA implementation. Furthermore, the scheme can
easily be implemented in any existing GKBA code without internal
modifications. The NEGF--GKBA+IC equation widens the class of
nonequilibrium phenomena considered so far, allowing for larger systems and/or longer time
propagations than was previously feasible. We also emphasize that the
proposed scheme
is compatible with any technique to obtain the initially correlated
density matrix as it does
not rely on the adiabatic switching procedure. In fact,
the NEGF--GKBA+IC equation is also suitable  to study systems at finite
temperature (the adiabatic switching procedure is consistent
only at zero temperature).

An interesting future prospect is the implementation of many-body  approximations
to the correlation self-energy that go beyond the ones currently used within the
GKBA. We derived a feasible form for the IC integral in the 2B
approximation, but the fundamental idea is completely general.
Indeed, in Appendix~\ref{appendixA} we provide an expression of the
IC integral for the GW$^{\eq}$ approximation, where the dynamically
screened interaction is taken from an equilibrium calculation.
For other commonly used many-body approximations, like
the full GW and T-matrix approximation, it is first necessary to find a
GKBA-like form of the screened interaction $W$ and T-matrix $T$ for
otherwise the favourable quadratic scaling with the number of time
steps is lost. Perhaps a more immediate direction is the application of the
NEGF--GKBA+IC scheme to open systems. This would allow for more
efficiently studying, for example, transient quantum transport or
photoionization in molecules.

\begin{acknowledgments}
D.K. acknowledges the Academy of Finland for funding under Project
No. 308697. G.S. and E.P.  acknowledge EC funding through the RISE Co-ExAN (Grant No. GA644076).
E.P. also acknowledges funding from the European Union project
MaX Materials design at the eXascale H2020-EINFRA-2015-1, Grant Agreement No.
676598 and Nanoscience Foundries and
Fine Analysis-Europe H2020-INFRAIA-2014-2015, Grant Agreement No. 654360.
\end{acknowledgments}

\appendix

\section{Generalized fluctuation-dissipation theorem}
\label{generalizedFD}
The purpose of this appendix is to present a generalized version of the fluctuation-dissipation theorem for the nonequilibrium Green's function, which will be used to derive the equivalence between the two different expressions for the IC integral, \Eq{ICIntegral} and \Eq{CollisionIntegralInit}.

Without any loss of generality, we consider a system in thermal equilibrium at inverse temperature $\beta$ and chemical potential $\mu$ for times $t\leq 0$, and out of equilibrium for $t>0$. Let $\hat{\varrho}$ be the many-body thermal density matrix with partition function $\mathcal{Z}$:
\begin{equation}
 \hat{\varrho} = \frac{e^{-\beta(\Hh - \mu \Nh)}}{\mathcal{Z}}
 = \sum_k \varrho_k |\psi_k \rangle \langle \psi_k |,
\end{equation}
where $|\psi_k \rangle$ are the many-body eigenstates of $\Hh$ with eigenvalue $E_k$ and number of particles $N_k$. Then,
\begin{equation}
 \varrho_k = \frac{e^{-\beta(E_k - \mu N_k)}}{\mathcal{Z}}. \label{thermalDensMat}
\end{equation}

By definition,
the exact lesser and greater Green's function read~\cite{Stefanucci2013}
\begin{align}
 \G^>_{ji} (t,t') = -i \sum_k \varrho_k \langle \psi_k | \dh_{j,H}(t)
 \dhd_{i,H}(t') | \psi_k \rangle ,
 \label{defg>}
 \\
 \G^<_{ji} (t,t') = i \sum_k \varrho_k
 \langle \psi_k | \dhd_{i,H}(t') \dh_{j,H}(t) | \psi_k \rangle,
 \label{defg<}
\end{align}
where the subscript $H$ denotes operators in the Heisenberg picture.
Taking into account that the system is in equilibrium for $t'< 0$
we have
%
% For times $t'<t_0$, the system is in equilibrium. As such, we can write
\begin{equation}
 \dh^{(\dagger)}_{i,H}(t') = %e^{i\Hh(t'-t_0)} \dhd_i e^{-i\Hh(t'-t_0)},
  e^{i\Hh t'} \dh^{(\dagger)}_i e^{-i\Hh t'},
\end{equation}
where $\dh^{(\dagger)}_i$ is the annihilation (creation) operator in
the Schr\"odinger picture. Thus, for  $t'< 0$, the lesser/greater Green's function can  be written as
\begin{align}
 \G^>_{ji} (t,t')
&= -i \sum_{kp} \varrho_k
 \langle \psi_k | \dh_{j,H}(t)
 |\psi_p \rangle \langle \psi_p | \dhd_i
  | \psi_k \rangle
\nonumber \\ &\times e^{i(E_p-E_k) t'} ,
 \label{G>1left}
\end{align}
\begin{align}
 \G^<_{ji} (t,t')
&= i \sum_{kp} \varrho_k
 \langle \psi_k | \dhd_i
 |\psi_p \rangle \langle \psi_p |
 \dh_{j,H}(t) | \psi_k \rangle
\nonumber \\ &\times  e^{i(E_k-E_p) t'},
 \label{G<1left}
\end{align}
where we inserted the completeness relation $\oneh = \sum_p |\psi_p
\rangle \langle \psi_p |$ between the fermionic operators.

Although the above expressions yield $\G^\lessgtr$ only for $t'<0$,
the right hand sides are well defined for all $t'$. We then define
the {\em auxiliary} lesser and greater Green's functions
$\G^{\lessgtr}_{{\rm aux1}}(t,t')$ as the right hand sides of
Eqs.~(\ref{G>1left},\ref{G<1left}) for all $t$ and $t'$.
Naturally, $\G^{\lessgtr}_{{\rm aux1}}(t,t') = \G^\lessgtr(t,t')$ only for $t'<0$.
We now show that these auxiliary functions satisfy a
generalization of the fluctuation-dissipation theorem.

Let us consider the Fourier transform of the auxiliary functions:
\begin{align}
 \G_{{\rm aux1}}^\lessgtr(t,t') =
 \int \frac{\dif \omega}{2 \pi} e^{i\omega t'} \G^\lessgtr_{{\rm aux1}} (t,\omega).
 \label{GFourier}
\end{align}
From Eqs.~(\ref{G>1left},\ref{G<1left}) it is straighforward to get
\begin{align}
\G_{{\rm aux1},ji}^{>} (t,\omega) &=
-2 \pi i \sum_{kp} \varrho_k
\langle \psi_k | \dh_{j,H}(t)
|\psi_p \rangle \langle \psi_p | \dhd_i
| \psi_k \rangle
\nonumber \\ &\times \delta(\omega - E_p + E_k ), \label{G>delta}
\end{align}
\begin{align}
\G_{{\rm aux1},ji}^{<} (t,\omega) &=
2\pi i \sum_{kp} \varrho_k
\langle \psi_k | \dhd_i
|\psi_p \rangle \langle \psi_p |
\dh_{j,H}(t) | \psi_k \rangle
\nonumber \\ &\times  \delta(\omega - E_k + E_p ) .\label{G<delta}
\end{align}
Inserting  the obvious relation
\begin{equation}
 \varrho_k = e^{-\beta(E_k-E_p) + \beta \mu(N_k-N_p)} \varrho_p, \label{thermalDensMatRelation}
\end{equation}
in \Eq{G>delta}, taking into account that only states fulfilling
$N_k=N_p-1$ contribute, and renaming the indices $k\leftrightarrow p$  we obtain
\begin{equation}
\G_{{\rm aux1}}^{<}(t,\omega) = -e^{-\beta (\omega - \mu)}
\G_{{\rm aux1}}^{>}(t,\omega).
\label{G><1}
\end{equation}
The lesser and greater auxiliary functions can be used to define the
retarded and advanced auxiliary functions in the usual manner
\begin{align*}
\G_{{\rm aux1}}^{R/A}(t,t') =
\mp i \theta \left (\mp t\pm t'\right) \left [ \G_{{\rm
aux1}}^>(t,t') - \G_{{\rm aux1}}^<(t,t') \right].
\end{align*}
It is straightfoward to verify that
$\G_{{\rm aux1}}^{>} (t,t') - \G_{{\rm aux1}}^<(t,t') =
\G_{{\rm aux1}}^R(t,t') - \G_{{\rm aux1}}^A(t,t')$.
Fourier transforming the retarded and advanced functions as in \Eq{GFourier} and taking into account
\Eq{G><1} we find a generalization of the fluctuation-dissipation
relations
\begin{align}
\begin{split}
\G_{{\rm aux1}}^> (t,\omega) &=
\bar{f}(\omega-\mu)
\left [\G_{{\rm aux1}}^R(t,\omega) - \G_{{\rm aux1}}^A(t,\omega)
\right ],
\\
\G_{{\rm aux1}}^<(t,\omega) & =
-f(\omega-\mu) \left [\G_{{\rm aux1}}^R(t,\omega) - \G_{{\rm
aux1}}^A(t,\omega) \right ] ,
\label{FDrel1}
\end{split}
\end{align}
with Fermi function
\begin{equation}
 f(\omega) = \frac{1}{e^{\beta \omega} +1}, \quad \text{and} \quad \bar{f}(\omega) = 1-f(\omega).
\end{equation}

A generalized version of the  fluctuation-dissipation
relations exists also for $t<0$ and $t'$ arbitrary. In this case,
from Eqs.~(\ref{defg>},\ref{defg<}) we find
\begin{align}
  \G^>_{ji} (t,t') &= -i \sum_{kp} \varrho_k \langle \psi_k | \dh_j
  |\psi_p \rangle \langle \psi_p |
  \dhd_{i,H}(t') | \psi_k\rangle
  \nonumber \\
  &\times  e^{i(E_k-E_p) t}, \label{G>eq}
\end{align}
\begin{align}
 \G^<_{ji} (t,t') &= i \sum_{kp} \varrho_k
 \langle \psi_k | \dhd_{i,H}(t')
 |\psi_p \rangle \langle \psi_p |
 \dh_j  | \psi_k \rangle
  \nonumber \\
  &\times e^{i(E_p-E_k) t}.
\end{align}
Again the functions on the right hand sides are well defined for all
$t$ and $t'$ and we denote them by $\G_{{\rm aux2}}^\lessgtr(t,t')$.
Of course,  $\G_{{\rm aux2}}^\lessgtr(t,t')=\G^\lessgtr(t,t')$
only  for $t<0$. A derivation similar to the one presented for $\G_{{\rm
aux1}}$ can be carried out for this other type of auxiliary functions
leading to
\begin{align}
\begin{split}
 \G_{{\rm aux2}}^>(\omega,t')
 &=
 \bar{f}(\omega-\mu) \left [ \G_{{\rm aux2}}^R(\omega,t') - \G_{{\rm
 aux2}}^A(\omega,t') \right ],
 \\
 \G_{{\rm aux2}}^<(\omega,t') &=
 -f(\omega-\mu)\left [\G_{{\rm aux2}}^R(\omega,t') - \G_{{\rm
 aux2}}^A(\omega,t') \right ],\label{FDrel2}
 \end{split}
\end{align}
where we defined
\begin{align}
 \G_{{\rm aux2}}^\lessgtr(t,t') =
 \int \frac{\dif \omega}{2 \pi} e^{-i\omega t} \G^\lessgtr_{{\rm aux2}} (\omega,t').
 \label{GFourier2}
\end{align}

In the derivation of the generalized fluctuation-dissipation
relations, \Eq{FDrel1} and \Eq{FDrel2}, the only property of the
operators $\dh_j$ that we have explicitly used is that
its action on a state  with $N$ particles
yields a state with $N-1$ particles. Under the same considerations as in Ref.~\cite{Stefanucci2013}, this leads to a generalized fluctuation-dissipation
relation also for the correlation self-energy, since
\begin{align}
 \Sigma^>_{ji} (t,t') = -i \trace{\hat{\varrho}\, \gammah_{j,H}(t)
 \gammah^\dagger_{i,H}(t')}_{\rm{irr}}, \\
 \Sigma^<_{ji} (t,t') = i \trace{\hat{\varrho}\,
 \gammah^\dagger_{i,H}(t') \gammah_{j,H}(t)}_{\rm{irr}},\label{S<}
\end{align}
where the operators $\gammah_i \equiv \sum_{mnp} v_{imnp} \dhd_m \dh_n \dh_p$ and the subscript ``$\rm{irr}$'' denotes the irreducible part of the correlator~\cite{Danielewicz1984,Stefanucci2013}. Using the same notation as for the auxiliary Green's functions we then have
\begin{align}
\begin{split}
  \S_{\rm aux1}^> (t,\omega) &=
  \bar{f}(\omega-\mu)
  \left [\S_{\rm aux1}^R(t,\omega) - \S_{\rm aux1}^A(t,\omega) \right
  ],
  \\
    \S_{\rm aux1}^<(t,\omega) & =
   -f(\omega-\mu) \left [\S_{\rm aux1}^R(t,\omega) - \S_{\rm
   aux1}^A(t,\omega) \right ],
\end{split}\label{FDSrel2}
   \end{align}
and similarly
\begin{align}
\begin{split}
  \S_{\rm aux2}^>(\omega,t')
  &=
 \bar{f}(\omega-\mu) \left [ \S_{\rm aux2}^R(\omega,t') - \S_{\rm
 aux2}^A(\omega,t') \right ],
 \\
 \S_{\rm aux2}^<(\omega,t') &=
 -f(\omega-\mu)\left [\S_{\rm aux2}^R(\omega,t') - \S_{\rm
 aux2}^A(\omega,t') \right ].
 \end{split} \label{FDSrel1}
\end{align}

We emphasize that \Eq{FDrel1}, \Eq{FDrel2}, \Eq{FDSrel2} and \Eq{FDSrel1} are {\em exact} relations. In the next appendix we use them to obtain an equivalent expression of the IC integral.

\section{IC integral in terms of real-time lesser and greater Green's functions}
\label{iceqformapp}

We consider the time-off-diagonal generalization of the IC integral in \Eq{ICIntegral}:
\begin{align}
\begin{split}
I^\ic(t_1,t_2) &=
 -i \int _0 ^\beta \dif \tau \Sigmacal^\rceil(t_1,\tau) \G^\lceil(\tau,t_2). \label{genICIntegral}
%  \\
% & = -i \int_0^\beta \dif \tau \Sigmacal^<(t_1,-i\tau) \G^>(-i\tau,t_2),
\end{split}
 \end{align}
When setting $t_{1}=t_{2}=t>0$, we obtain the original IC integral, i.e. $\Icalh^\ic(t)=I^\ic(t,t)$. To make use of the generalized fluctuation-dissipation theorems in \Eq{genICIntegral}, we rewrite $\S^\rceil$ and $\G^\lceil$ in terms of $\S^<$ and $\G^>$, respectively. This rewriting can be done by the same considerations as in Ref.~\cite{Stefanucci2013}, and reads
\begin{align*}
 \G^\lceil(\tau,t_2) = e^{\mu \tau} \G^> (-i\tau,t_2), \\
\S^\rceil(t_1,\tau) = e^{-\mu \tau} \S^< (t_1,-i\tau),
\end{align*}
where the lesser and greater functions for complex times are defined via the analytic continuation $t \to -i\tau$ in \Eq{G>eq} and \Eq{S<}. Moreover, these lesser and greater functions coincide with their respective auxiliary quantities since the first (second) argument is governed by the equilibrium Hamiltonian. The factors containing $\mu$ in the above equations will cancel when inserted into \Eq{genICIntegral}, yielding
\begin{align}
I^\ic(t_1,t_2) =
-i \int_0^\beta \dif \tau \S^< _{\rm aux1}(t_1,-i\tau) \G^>_{\rm aux2}(-i\tau,t_2).
 \end{align}

Furthermore, the Fourier transform of the auxiliary quantities, \Eq{GFourier} and \Eq{GFourier2}, yields
\begin{align*}
\G^>_{\rm aux2}(-i\tau,t_2) = \int \frac{\dif \omega_2}{2\pi} e^{-i\omega_2 (-i\tau)} \G^>_{\rm aux2}(\omega_2,t_2) \\
 \S^<_{\rm aux1}(t_1,-i\tau) = \int \frac{\dif \omega_1}{2\pi} e^{i\omega_1 (-i\tau)} \S^< _{\rm aux1}(t_1,\omega_1).
\end{align*}
Inserting these relations into \Eq{genICIntegral} and integrating over $\tau$ yields
\begin{align}
I^\ic(t_1,t_2)
= -i \int & \frac{\dif \omega_1 \dif \omega_2}{4\pi^2}
\frac{e^{ \beta(\omega_1-\omega_2)}-1}{\omega_1-\omega_2-i\eta}
\nonumber \\
&\times \S_{\rm aux1}^<(t_1,\omega_1) \G_{\rm aux2}^>(\omega_2,t_2),
\end{align}
where we, in the denominator, added to $\omega_1$ a small negative imaginary part
$\omega_1 \to \omega_1 - i \eta/2$, and likewise $\omega_2 \to \omega_2 + i \eta/2$
to regularize the integral (the limit $\eta\to 0$ should be taken at the end of the
calculation~\cite{Stefanucci2013}).

We now use the generalized fluctuation-dissipation relations for
$\S_{\rm aux1}^<$ and $\G_{\rm aux2}^>$ which, together with the
relation
\begin{equation*}
 f(\omega_1-\mu) \bar{f}(\omega_2-\mu) \left (
 e^{\beta(\omega_1-\omega_2)}-1 \right) = f(\omega_2-\mu) - f(\omega_1-\mu),
\end{equation*}
yields
\begin{widetext}
\begin{align}
I^\ic(t_1,t_2)=-i\int
\frac{\dif \omega_1 \dif \omega_2}{4\pi^2}
\frac{\left [\S_{\rm aux1}^R(t_1,\omega_1) - \S_{\rm
aux1}^A(t_1,\omega_1) \right ]
\G_{\rm aux2}^<(\omega_2,t_2)
-
\S_{\rm aux1}^<(t_1,\omega_1)
\left [ \G_{\rm aux2}^R(\omega_2,t_2) - \G_{\rm aux2}^A(\omega_2,t_2)
\right ]}{\omega_1-\omega_2-i\eta}.
\label{icstep3}
\end{align}
Writing the denominator as
 \begin{equation}
 \frac{1}{i(\omega_1-\omega_2-i\eta)} =
 \int_{-\infty}^0 \dif \tb\,  e^{i(\omega_1-\omega_2-i\eta)\tb},
 \end{equation}
and recognizing the inverse Fourier transform of the quantities under
the integral sign in \Eq{icstep3} we can write
\begin{align}
	I^\ic(t_1,t_2) =
\int_{-\infty}^0 \!\!\dif
\tb
\left [
\left (\S_{\rm aux1}^R(t_1,\tb) - \S_{\rm aux1}^A(t_1,\tb) \right )
\G_{\rm aux2}^<(\tb,t_2)
-\S_{\rm aux1}^<(t_1,\tb)
\left ( \G_{\rm aux2}^R(\tb,t_2) - \G_{\rm aux2}^A(\tb,t_2) \right )
\right ]e^{\eta \tb}.
\label{timeResultAppendix}
\end{align}
Next we observe that for $\tb < 0$ the first auxiliary self-energy $\S_{\rm aux1}$ is identical to the self-energy $\S$ and the second auxiliary Green's function $\G_{\rm aux2}$ is identical to the Green's function $\G$. Furthermore, using $\G^> - \G^< = \G^R - \G^A$, and likewise for $\S$, we find
%
% see that for $\tb < 0$ and $t_{1},t_{2}>0$ we have $\S^A(t_1,\tb)=\G^R(\tb,t_2)=0$ and therefore
\begin{align}
\begin{split}
I^\ic(t_1,t_2)=
   \int_{-\infty}^0 \dif \tb \
\left[
\S^>(t_1,\tb) \G^<(\tb,t_2)
-
\S^<(t_1,\tb)
\G^>(\tb,t_2)
\right]e^{\eta \tb}.
\end{split}\label{equivalentForm}
\end{align}
\end{widetext}
If we now let $t_2 \to t_1=t>0$, we obtain the sought-after equivalence between the two alternative forms of the IC integral, i.e., \Eq{ICIntegral} and \Eq{CollisionIntegralInit}.

\section{Initial correlations for the GW$^{\rm eq}$ approximation}
\label{appendixA}
Let us consider the GW approximation to the lesser and greater self-energy
\begin{equation}
 \Sigmacalh_{ij}^\lessgtr (t,\tb) = i \G^\lessgtr_{ij}(t,\tb) W^\lessgtr_{ij}(t,\tb), \label{GW}
\end{equation}
and let us insert the relation \Eq{GKBAapprox} in \Eq{GW}.
The first term $\Jcalh^{\ic}_{ik} (t)$ of the
IC integral in \Eq{CollisionIntegralInit} reads
\begin{align}
\begin{split}
 \Jcalh^{\ic}_{ik} (t) &=
 i \sum_{m n p} \Gcalh^R_{in}(0,t) \\
 &\times
 \left [ \int_{-\infty}^0 \!\!\dif \tb \ \Gcalh^>_{nm} (0,\tb) W^>_{im}(t,\tb) \Gcalh^<_{mp}(\tb, 0)
 \right ] \Gcalh^A_{pk}(0,t).
\end{split}
 \end{align}
In a full GW calculation
the screened interaction $W$ is a functional of the nonequilibrium
$\G$. In this case the use of the GKBA does not reduce
the cubic scaling with the number of time steps of the KBE.

To introduce dynamical screening to some extent we consider a \emph{prescribed} screened interaction like, e.g., the one resulting from an equilibrium calculation $W=W^{\rm eq}$. This is the case in $GW_0$, where $W_0$ is obtained from the RPA response function with bare or HF equilibrium Green's function. Another example is the $W$ of Refs.~\cite{Pal2009,Pal2011} where the response function satisfies the Bethe-Salpeter equation with a frequency-dependent kernel. The advantage of working with an equilibrium screened interaction is that $W^{\rm eq}(t,\tb)$ depends only on $t-\tb$. Fourier transforming according to
\begin{align}
W^{\rm eq,\lessgtr}(t,\tb) =
\int \frac{\dif \omega}{2\pi} e^{-i\omega(t-\tb)}
W^{\rm eq,\lessgtr}(\omega),
 \end{align}
we obtain
\begin{align}
 \Jcalh^{\ic}_{ik} (t) =
 i \sum_{n p} \Gcalh^R_{in}(0,t)
 f_{inp} (t) \Gcalh^A_{pk}(0,t),
 \end{align}
where the tensor $f_{inp} (t) = \int \frac{\dif \omega}{2\pi}
e^{-i\omega t} f_{inp} (\omega)$ is the Fourier transform of
\begin{equation*}
 f_{inp} (\omega) = \sum_m W^{\rm eq,>}_{im}(\omega) \int_{-\infty}^0
 \dif \tb \ e^{i\omega \tb} \Gcalh^>_{nm} (0,\tb)  \Gcalh^<_{mp}(\tb,
 0).
\end{equation*}
The time integral in the expression above can be done analytically by
using \Eq{GlesserGreaterSmallT} and one finds
\begin{equation}
  f_{inp} (\omega) = \frac{1}{i(\omega + \epsilon_n - \epsilon_p) +
  \eta} \sum_m \bar{\rho}^\eq_{nm} W^{\rm eq,>}_{im}(\omega) \rho^\eq_{mp}.
\end{equation}
The same considerations can be applied to the second term
$\bar{\Jcalh}^{\ic}_{ik} (t)$ of the IC integral in \Eq{CollisionIntegralInit},
yielding an expression for the IC integral suitable for numerical
implementations. In fact, the tensor $f_{inp}(t)$ can be calculated
separately for the needed time interval, and the
computational effort scale linearly with the number of time steps.
Furthermore, the IC integral in the GW$^{\eq}$ approximation scales
with the fourth power of the number of basis functions $N_{b}$
(this should be compared with the $N_{b}^{5}$ scaling of IC integral
in the 2B approximation). Thus, the GW$^{\eq}$ approximation
is numerically feasible.

% \bibliography{bibliography}
%merlin.mbs apsrev4-1.bst 2010-07-25 4.21a (PWD, AO, DPC) hacked
%Control: key (0)
%Control: author (8) initials jnrlst
%Control: editor formatted (1) identically to author
%Control: production of article title (-1) disabled
%Control: page (0) single
%Control: year (1) truncated
%Control: production of eprint (0) enabled
%
\end{document}